\documentclass[12pt,twoside]{iopart}

\usepackage[bottom]{footmisc} 
\DefineFNsymbols{rjnumber}{{*}{^1}{^2}{^3}{^4}{^5}{^6}{^7}{^8}{^9}
{^{10}}{^{11}}{^{12}}{^{13}}{^{14}}{^{15}}{^{16}}{^{17}}{^{18}}{^{19}}
{^{20}}{^{21}}{^{22}}{^{23}}{^{24}}{^{25}}{^{26}}{^{27}}{^{28}}{^{29}}
{^{30}}{^{31}}{^{32}}{^{33}}{^{34}}{^{35}}{^{36}}{^{37}}{^{38}}{^{39}}{^{40}}} 
\setfnsymbol{rjnumber} 
\usepackage{psfrag}
\usepackage{epsfig}
\newcommand{\eq}[1]{(\ref{#1})}

\newcommand{\fig}[1]{figure \ref{#1}}

\begin{document}

\title{Optical geometry across the horizon}%
\author{Rickard Jonsson\\[2mm]%
{\small \it Department of Theoretical Physics, Chalmers University of Technology, 41296 G\"oteborg,
Sweden.}\\[2mm]
\small{\rm E-mail: rico@fy.chalmers.se}
\\[2mm]
\small{\rm
Submitted 2004-12-10, Published 2005-12-08\\
	 Journal Reference: Class. Quantum Grav. {\bf 23} 77
}
}
\begin{abstract}%
In a companion paper (Jonsson and Westman 2006 {\it Class. Quantum Grav.} {\bf 23} 61), a generalization of optical
geometry, assuming a non-shearing reference congruence, is
discussed. 
Here we illustrate that this formalism can be applied to (a finite
four-volume) of any
spherically symmetric spacetime.  
In particular 
we apply the formalism, using a non-static
reference congruence, to do optical geometry across the horizon of a
static black hole. 
While the resulting geometry in principle is time
dependent, we can choose the reference congruence in such a manner
that an embedding of the geometry always looks the same. Relative to
the embedded geometry the reference points are then moving. We discuss the
motion of photons, inertial forces and gyroscope precession in this framework.
\end{abstract}
\setcounter{page}{1}
\section{Introduction}
In \cite{genopt} it is illustrated how we can generalize the optical
geometry (see e.g \cite{optiskintro} for an introduction) to a wider
class of spacetimes than the conformally static ones. In particular,
employing the standard projected curvature (see \cite{genopt} for
alternative curvature measures), the new class of
spacetimes consists of those spacetimes that admit a hypersurface forming shearfree congruence of timelike worldlines.
We are now
curious as to whether any of the standard solutions to Einstein's
equations, that are {\it not} conformally static, 
falls into the new category. The task is then to look for a
congruence such that, in the corresponding coordinates, the metric
after rescaling takes the form \cite{genopt}
\begin{equation} \label{finale}
\tilde{g}_{\mu\nu}=\left[
\begin{array}{lllll}
\ 1 \ \ \ \ \ \ \  \ \ \ 0  \ \ \ \ \ \ \  		\\[0mm]
\ 				\\[0mm]
\ 0 \ \ \ \ -e^{2 \Omega(t,\bf{x})} \bar{h}_{ij}({\bf x})
\end{array}
\right].
\end{equation}
Indeed, as will be shown in the coming section, such a congruence can
be found  (in a finite four-volume of the spacetime) whenever we have
spherical symmetry in the original metric. This includes the inside of
a Schwarzschild black hole and the horizon as well.

Throughout the article, we will use the timelike $(+,-,-,-)$
convention for the sign of the metric. Also the optical line element will be denoted by $d\tilde{s}$.

\section{A spherical line element}
A general, time dependent, spherically symmetric line element can be written on the form
\begin{eqnarray}
d\tau^2=a(r,t)dt^2-2b(r,t)dr dt - c(r,t)dr^2  -r^2d\Omega^2.
\end{eqnarray}
This we may rewrite as
\begin{eqnarray}
d\tau^2=r^2 \left(  d\bar{\tau}^2 -d\Omega^2 \right).
\end{eqnarray}
Here we have introduced a two-dimensional line element  
\begin{eqnarray}\label{redlin}
d\bar{\tau}^2=\frac{a(r,t)}{r^2} dt^2-  \frac{2b(r,t)}{r^2}dr dt - \frac{c(r,t)}{r^2} dr^2 .
\end{eqnarray}
In this reduced spacetime we may introduce an arbitrary timelike initial congruence line. From this line we go a proper orthogonal distance $ds$ to create a new line. From the new line we create yet another line in the same manner. Next we introduce a new spatial coordinate $x'$ that is constant for every congruence line, and where $dx'=ds$. Also introduce time slices of constant $t'$ orthogonal to the congruence lines. The reduced line element then takes the form
\begin{eqnarray}
d\bar{\tau}^2=f(t',x') dt'^2 - dx'^2.
\end{eqnarray} 
The full line element with respect to these coordinates is then on the form
\begin{eqnarray}
d\tau^2=r^2 f(t', x') \left(  dt'^2- \frac{1}{f(x',t')} \left(dx'^2  +d\Omega^2 \right) \right).
\end{eqnarray}
Here $r$ is in principle known in terms of $x'$ and $t'$. After
rescaling away the factor $r^2 f(t', x')$,  this line element clearly has the form of \eq{finale} needed for the generalized optical geometry. The optical geometry is then
\begin{eqnarray}
d\tilde{s}^2= \frac{1}{f(x',t')} \left(dx'^2  +d\Omega^2 \right).
\end{eqnarray}
So, in whatever spherically symmetric spacetime we consider we can thus do
the generalized optical geometry. This includes collapsing stars, the
spacetime around the horizon of a Schwarzschild black hole and so
forth. Notice however that there is no guarantee for the generalized
geometry to work {\it globally} in these spacetimes. The way we are
constructing our congruence, it may for instance go null before we
have come very far from our original congruence line. Also, the
geometry which will be determined by how we choose our initial
congruence line, may be more or less complicated, time dependent and
so forth. 

\subsection{A small note on intuition}
\begin{figure}[ht]
  \begin{center}
      	\epsfig{figure=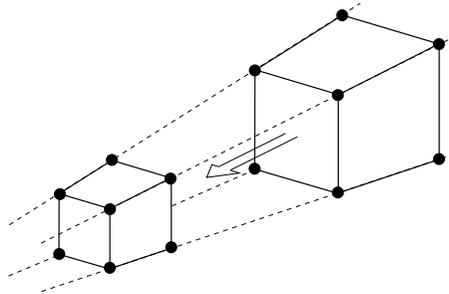,width=6cm,angle=0}
      	\caption{To have vanishing shear (which is necessary for a
      	congruence generating the optical geometry), the congruence points (the black dots), must be
      	shifted in such a way that the shape of a little box of
      	congruence points (as seen
      	when comoving with the box) is the same at all times.}  
     	\label{noshearx}
  \end{center} 
\end{figure}

We are here considering a congruence that is fixed in the spherical
angles. From a dynamical point of view, we have found a radial
velocity of infinitesimally separated congruence points such that the
proper {\it shape} that is spanned by the points is preserved,
see \fig{noshearx}. 

If we start with a cube it must remain a cube, but not necessarily of
the same size, in a system comoving with the points. For instance
considering a flat space and low velocities inwards (towards the
origin of the spherical coordinates) we may understand that the
velocity of the inner part of the cube must be smaller than that of
the outer part to insure that the cube is not elongated in the radial direction.

\section{Optical geometry for a black hole including the horizon}
Let us now study a black hole explicitly, with focus on the
horizon. The ordinary Schwarzschild coordinates are ill suited for
congruences passing through the horizon. There is however another coordinate
system (called Painlev\'e coordinates) in which the Schwarzschild line element is given by
\begin{eqnarray}
d\tau^2=\left( 1 - \frac{2M}{r}  \right) dT_{\textrm{\tiny P}}^2-2\sqrt{\frac{2M}{r}} dT_{\textrm{\tiny P}}dr-dr^2 -r^2 d\Omega^2.
\end{eqnarray}
This line element is connected to the standard line element through a
resetting of the ordinary Schwarzschild clocks\footnote{The clocks are
reset in such a way that the coordinate time passed for a freely
falling observer initially at rest at infinity corresponds to the
proper time experienced by this observer. Inside the horizon one cannot
have any material clocks at a fixed $r$ but that doesn't matter.}. 
In these coordinates there are no problems in passing through the horizon.
We also find it practical to introduce dimensionless coordinates $r/2M \rightarrow x$, $T_{\textrm{\tiny P}}/2M \rightarrow T$. Then the line element takes the form
\begin{eqnarray}
\frac{d\tau^2}{(2M)^2}=\left( 1 - \frac{1}{x}  \right) dT^2-2\frac{1}{\sqrt{x}} dTdx-dx^2 -x^2 d\Omega^2.
\end{eqnarray}
The reduced line element (compare with \eq{redlin}) is then given by
\begin{eqnarray}\label{gali}
d\bar{\tau}^2=A dT^2 - 2 B dT dx -C dx^2.
\end{eqnarray}
Here the reduced metrical components are given by%
\footnote{
One may alternatively use the Eddington Finkelstein coordinates where after corresponding rescalings $A=(1-1/x)/x^2$,  $B=1/x^2$, $C=0$.}
\begin{eqnarray}\label{abc}
A=\frac{1}{x^2}\left(1-\frac{1}{x}\right) \quad B=\frac{1}{x^2}\frac{1}{\sqrt{x}} \quad  C=\frac{1}{x^2} .
\end{eqnarray}
Now we may introduce an arbitrary timelike trajectory that passes the
horizon. From this we go a proper distance $ds$ orthogonal to the
trajectory, to create a new congruence line and so forth. In general
this scheme would completely hide the manifest time symmetry of the black hole. There is however a way to circumvent this as will be shown in the following sections.

\section{Keeping the time symmetry, covariant approach}
\begin{figure}[ht]
  \begin{center}
	\psfrag{x}{$x$}
	\psfrag{T}{$\hspace{-0.5mm}T$}
	\psfrag{D}{$\Delta T$}
	\psfrag{s}{\tiny $\Delta s$}
	\epsfig{figure=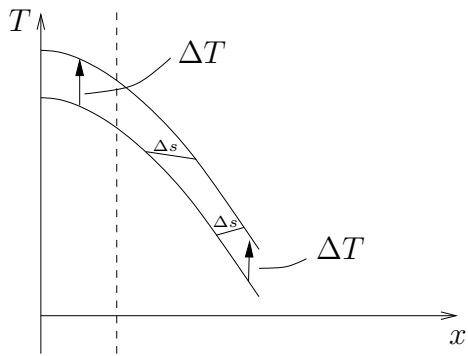,width=6cm,angle=0}
      	\caption{Two congruence lines separated along the Killing field with constant proper distance between them.}  
     	\label{haa}
  \end{center} 
\end{figure}

Suppose that we can find an initial trajectory such that the second
trajectory (go $ds$ orthogonal to the initial trajectory) is related to the first by a simple translation straight in the $T$-direction (along the Killing field connected to $T$). This way we would maintain a certain time symmetry in the optical metric. In \fig{haa} we see schematically how this would work.

\begin{figure}[ht]
  \begin{center}
	\psfrag{K}{$K \hspace{0.5mm} ds \hspace{0.5mm} \xi^\mu$}
	\psfrag{d}{$ds \hspace{0.5mm} v^\mu$}
	\psfrag{s}{$\sigma \hspace{0.5mm} ds \hspace{0.5mm}  u^\mu$}
      	\epsfig{figure=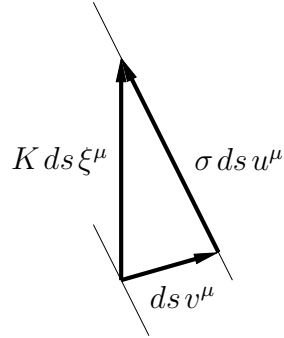,width=3cm,angle=0}
	\caption{The relation between Killing field $\xi^\mu$,
      	four-velocity $u^\mu$ and the orthogonal vector
      	$v^\mu$, assuming an in-going congruence.
      	}  
     	\label{cov}
  \end{center} 
\end{figure}

Zooming in on the two lines around some specific point, they will to
first order be two parallel straight lines. Given the tilt of the
lines (i.e. the four-velocity) we can find a relation between the
displacement along the Killing field and the orthogonal distance
between the lines. How these are related is sketched in Fig
{\ref{cov}}. Here $u^\mu$ is the four-velocity of the trajectories and $v^\mu$ is a spacelike vector normed to $-1$ and orthogonal to $u^\mu$.
Just adding vectors we find
\begin{eqnarray}\label{start}
K ds \xi^\mu=ds v^\mu + \sigma ds u^\mu.
\end{eqnarray}
Notice that $K$ is a assumed to be a constant unlike $\sigma$.
Multiplying both sides by $u_\mu$ we get
\begin{eqnarray}\label{sigma}
\sigma  = K  \xi^\mu u_\mu.
\end{eqnarray}
Inserting this back into \eq{start} we get
\begin{eqnarray} \label{hopp}
K  \xi^\mu= v^\mu + K  \xi^\alpha u_\alpha u^\mu .
\end{eqnarray}
Taking the absolute value of both sides yields shortly
\begin{eqnarray}
K^2=\frac{1}{(\xi^\alpha u_\alpha)^2 - \xi^\alpha \xi_\alpha}.
\end{eqnarray}
So $K$ is known given $u^\mu$. Solving for $\xi^\alpha u_\alpha$ yields
\begin{eqnarray}\label{king}
\xi^\alpha u_\alpha = \pm \sqrt{\xi^\alpha \xi_\alpha + \frac{1}{K^2}}.
\end{eqnarray}
Outside of the horizon $\xi^\mu$ is always timelike and thus the sign in front of the root will be positive%
\footnote{Assuming $u^\mu$ and $\xi^\mu$ to both be future directed.}
. Inside of the horizon, where $\xi^\mu$ is spacelike, we can have both signs
depending on $u^\mu$. On the horizon we however have $\xi^\mu
\xi_\mu=0$. Thus for finite $K$ we realize that we must have the
positive sign on the inside as well to get a continuous four-velocity
across the horizon. Using \eq{king} together with $u^\alpha
u_\alpha=1$, we can in principle solve for $u^\mu$ given $K$. The
equations are however second order, and there will be four different
solutions at every point (see section \ref{com} for intuition). There
is however a simple way to get first order equations.

Defining $v^\mu$ to be the orthonormal vector to $u^\mu$ that lies less than $180{}^\circ$ clockwise%
\footnote{In accordance to \fig{cov}, assuming positive values of $K$, $\sigma$ and $ds$.}
of $u^\mu$, we may write
\begin{equation}\label{hard1}
v^\mu=\frac{1}{\sqrt{g}} \epsilon^{\mu\nu} g_{\nu \rho} u^\rho \qquad \textrm{where} \qquad \epsilon^{\mu \nu}
=\left(
\begin{array}{lllll}
0 \ \ \textrm{-}1 \\[0mm]
1 \ \ \ 0	 \\[0mm]
\end{array}
\right).
\end{equation}
Here $g=-\textrm{Det}(g_{\mu\nu})$. Then we may rewrite (\ref{hopp}) into
\begin{eqnarray}\label{hard0}
K \xi^\mu =\frac{1}{\sqrt{g}} \epsilon^{\mu\nu} g_{\nu \rho}  u^\rho  \pm K \sqrt{\xi^\alpha \xi_\alpha + \frac{1}{K^2}}  u^\mu.
\end{eqnarray}
As noted before, considering an in-falling congruence at the horizon
we should choose the positive sign. We see that \eq{hard0} is a linear equation system which we, given $K$, should be able
to solve to find $u^\mu$. The scheme thus appears successful and there
exists an optical (shearfree) congruence that will preserve manifest time symmetry.

\section{The congruence for a Schwarzschild black hole}
Choosing the positive sign of (\ref{hard0}) and assuming a positive $K$ in accordance with the discussion above, (\ref{hard0}) takes the form
\begin{eqnarray}\label{hard00}
K \xi^\mu =\frac{1}{\sqrt{g}} \epsilon^{\mu\nu} g_{\nu \rho}  u^\rho  + K \sqrt{\xi^\alpha \xi_\alpha + \frac{1}{K^2}}  u^\mu.
\end{eqnarray}
Assuming the reduced line element to be of the form of (\ref{gali}), and $\xi^\mu=(1,0)$ this can be written
\begin{eqnarray}\label{hard2}
K&=&\frac{B}{\sqrt{g}} u^0+\frac{C}{\sqrt{g}} u^1 +  \sqrt{ K^2 A + 1} u^0 \\
0 &=& \frac{1}{\sqrt{g}} (A u^0 - B u^1) {+} \sqrt{ K^2 A + 1}  u^1 . \label{hardadd}
\end{eqnarray}
Recognizing that $dx/dT=u^1/u^0$ we find from the second equation alone that

\begin{eqnarray}\label{aaaa}
\frac{dx}{dT}=\frac{A}{B - \sqrt{g} \sqrt{K^2 A +1}}.
\end{eqnarray}
Inserting the metrical components of (\ref{abc}) into (\ref{aaaa}) we find
\begin{eqnarray}\label{jo}
\frac{dT}{dx}=\frac{\frac{1}{\sqrt{x}}  - \sqrt{\frac{K^2}{x^2}(1-\frac{1}{x})  +1 }   }{1-\frac{1}{x}}.
\end{eqnarray}
So here we have the tilt of the reference congruence lines in the
Painlev\'e coordinates.
For an infinite value of the free parameter $K$, outside the horizon,
this corresponds to a congruence at rest (i.e. the classical optical congruence). Inside the horizon the root takes a negative value for infinite $K$ and we have no solution.

For any finite values of $K$ we notice that at infinity \eq{jo} will
correspond to an in-going photon ($\frac{dT}{dx}=-1$). In the particular case of $K=0$ the congruence will correspond to an in-going photon all the way through the horizon and into the singularity. Photons are however at first sight not particularly well suited for a congruence. For finite $K$ and $x<1$ we see that the root goes imaginary unless
\begin{eqnarray}
K^2 < \frac{x^3}{1-x}.
\end{eqnarray}
So the conclusion is that we can do optical geometry, while keeping manifest time symmetry, from a point arbitrarily close to the singularity, across the horizon and all the way towards infinity. Notice in particular that with this scheme we not only get the time symmetry, we insure that we can span the full spacetime all the way towards infinite Schwarzschild times.

\subsection{Comments}\label{com}
A short comment may be in order regarding the congruence as we approach infinity, where the spacetime approaches  Minkowski. Here one would expect that any congruence with fixed coordinate velocity $dx/dt$ would work as an optical congruence, not just left-moving photons. Indeed from (\ref{jo}) we see that for any large, but finite, $x$ there exists a $K$ such that we can have any in-going coordinate velocity of the congruence. As we go outwards towards infinity from this point the congruence will however start approaching a left moving photon.

\begin{figure}[ht]
  \begin{center}
      	\epsfig{figure=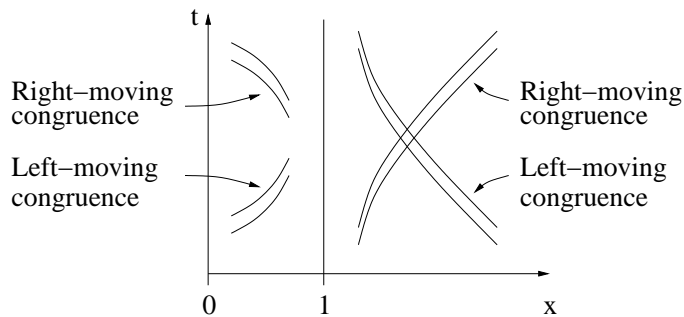,width=9cm,angle=0}
      	\caption{Illustrating in standard (dimensionless) Schwarzschild coordinates that for any left-moving congruence that
      	fulfills the requirements there is also a right-moving
      	congruence that fulfills the requirements.}  
     	\label{leftright}
  \end{center} 
\end{figure}

Another comment may be in order. The existence of an optical congruence, keeping manifest time symmetry, is independent of what coordinates we are using. In the standard Schwarzschild coordinates it is easy to realize that, both on the inside and the outside, the existence of {\it one} congruence immediately implies the existence of {\it another}%
\footnote{Except if the congruence on the outside would be parallel to the Killing field, or equivalently  if the congruence on the inside would be perpendicular to the Killing field.}
, as depicted in \fig{leftright}.

Inside the horizon this is manifesting itself in the $\pm$ sign of
(\ref{hard0}). On the outside, where we must have a plus in the $\pm$
sign, it manifests itself in the possibility to have negative $K$. In
the latter case we need to consider a slightly different image than
that of \fig{cov}, but the mathematics will be identical if we let $K$ assume negative values.  
In general we may show that
\begin{eqnarray}\label{aaaab}
\frac{dx}{dT}=\frac{A}{B \mp \sqrt{g} \sqrt{K^2 A +1}}.
\end{eqnarray}
The minus in this case corresponds to a left-moving congruence, and the plus a right-moving.

\section{The optical metric}
\begin{figure}[ht]
  \begin{center}
	\psfrag{K}{$K \hspace{0.5mm} ds \hspace{0.5mm} \xi^\mu$}
	\psfrag{d}{$ds \hspace{0.5mm} v^\mu$}
	\psfrag{s}{$\sigma \hspace{0.5mm} ds \hspace{0.5mm}  u^\mu$}
	\psfrag{t'+dt'}{$t'+dt'$}
	\psfrag{dT}{$dT$}
	\psfrag{t'}{$t'$}
	\epsfig{figure=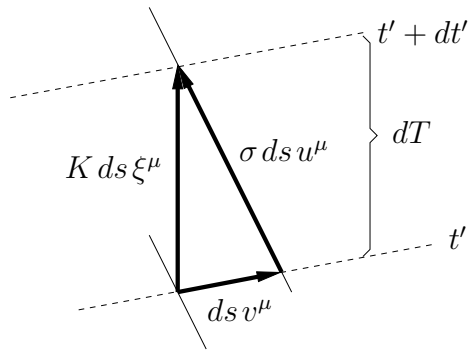,width=6cm,angle=0}
      	\caption{The relation between Killing field, four-velocity and the orthogonal vector. The dotted lines are the new local time slices}  
     	\label{cov2x}
  \end{center} 
\end{figure}

Let the spatial coordinate difference $dx'$, separating two nearby congruence lines, equal the proper orthogonal distance $ds$ between the lines%
\footnote{Recall that the distance between the lines is by definition
constant along the lines.}.
Let the new coordinate time difference $dt'$, separating two time slices, equal the original coordinate difference $dT$, as measured along the Killing field.
The reduced metric takes a new form according to
\begin{eqnarray}
d\bar{\tau}^2=A dT^2 - 2 B dT dx -C dx^2  \qquad \rightarrow \qquad d\bar{\tau}^2= f(x',t') dt'^2 - dx'^2.
\end{eqnarray}
Here $f$ is a function yet to be determined. 
Recall the relation between the various vectors, as depicted in \fig{cov2x}.

\begin{figure}[ht]
  \begin{center}
      	\epsfig{figure=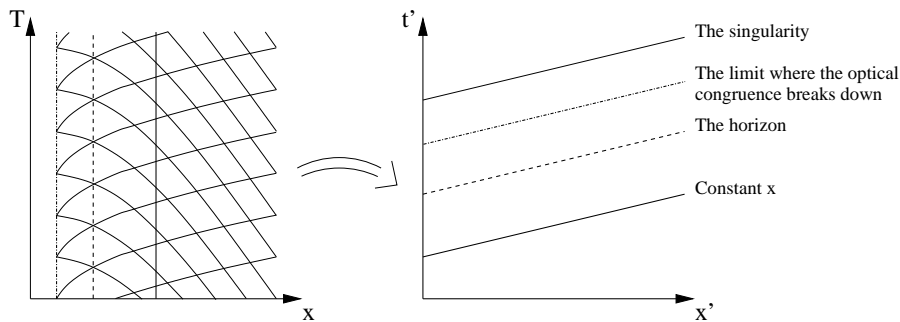,width=12cm,angle=0}
      	\caption{To the left the congruence in the Painlev\'e
      	coordinates. To the right the horizon etc relative to coordinates
      	adapted to the congruence.}  
     	\label{new}
  \end{center} 
\end{figure}

Like before we assume the Killing field to be ($1,0$) so that $dT=K ds=dt'$. 
The proper distance squared, as measured along a congruence line, separating two time slices can be expressed as
\begin{eqnarray}
d\bar{\tau}^2=(\sigma ds)^2 u^\mu u_\mu              \qquad     d\bar{\tau}^2= f dt'^2     .
\end{eqnarray}
From (\ref{sigma}) and (\ref{king}) respectively we have
\begin{eqnarray}
\sigma = K \xi^\alpha u_\alpha   \qquad  \quad  \xi^\alpha u_\alpha =  \sqrt{A + \frac{1}{K^2}}.
\end{eqnarray}
Like before we have chosen the positive sign of the root. Putting the pieces together we find
\begin{eqnarray}\label{bob}
f=A+\frac{1}{K^2}.
\end{eqnarray}
The total, original, line element in the new coordinates is now given by
\begin{eqnarray}
d\tau^2 =(2M)^2 x^2\left(      f(x) dt'^2 - dx'^2  -   d\Omega^2  \right) .
\end{eqnarray}
Using \eq{abc} and \eq{bob}, the optical metric is thus given by:
 \begin{eqnarray}\label{opmet}
d\tilde{s}^2=\frac{1}{   \frac{1}{x^2} \left( 1 - \frac{1}{x} \right) + \frac{1}{K^2}    } \left( dx'^2  +  d\Omega^2 \right).
\end{eqnarray} 
Notice however that it is not in explicit form since we do not have
$x$ in terms of $x'$ and $t'$. We may however recall \fig{cov2x}
where we for constant $x$ have $dt'=K ds= K dx'$ (since $dx'$ per
definition equals $ds$). Thus we know that constant $x$ means $dt'/dx'=K$. In the new coordinates we have therefore a schematic picture as depicted in \fig{new}.

Notice that the Killing field is still a constant vector (tilted up
and to the right) in the new coordinates. While we still do not have $x$ analytically in terms of $x'$ and $t'$, we know the qualitative relation well enough to understand some basic features.

\subsection{The rubber sheet model}
We see from \eq{opmet} that at spatial infinity the geometry becomes
that of a three-cylinder, except if $K$ is infinite. Also we see that
on the inner boundary, where the optical congruence breaks down, the
stretching (of $d\tilde{s}$ relative to $ds$) is infinite.

It appears very difficult to do any calculations in our new
coordinates, considering that we don't have any explicit relation for
$x$ in terms of $x'$ and $t'$. At every fixed $t'$ we may however
express the optical geometry as a function of $x$, given that we have
a relation between $dx$ and $dx'$ (as will be derived in section \ref{roref}). This background
geometry is time independent. The scenario (in 2D) can then be exactly
described by a rubber sheet sliding snuggly over the fixed background
geometry. Photons move on geodesics with unit velocity at every point
if we {\it comove} with the rubber sheet. The velocity of the rubber
sheet will correspond to the velocity of a constant $x$ line relative
to the reference congruence. Then we can use our knowledge of geodesics on rotational surfaces, and relative velocities, to find the paths of photons relative this pseudo-optical background geometry.

\section{On the relation between $x$ and $x'$}\label{roref}
Given a displacement $dx'$ along the $x'$-axis we want to find $dx$. From \fig{cov} we see that $dx=v^x ds$ or equivalently
\begin{eqnarray}\label{k1}
dx= v^x dx'.
\end{eqnarray}
From (\ref{hard1}) and (\ref{hard2}) we readily find
\begin{eqnarray}\label{k2}
v^x=-{\alpha} u^x    \quad \textrm{where} \quad {\alpha}=\sqrt{K^2 A +1}.
\end{eqnarray}
To find $u^x$, we solve the linear equation system of 
\eq{hard2} and \eq{hardadd}
letting $u^0 \rightarrow  u^T$ and $u^1 \rightarrow u^x$. 
The result is
\begin{eqnarray}
u^x=\frac{K}{ \left(  \frac{B^2}{\sqrt{g}^2} -(K^2 A +1)    \right)  \frac{\sqrt{g}}{A} + \frac{C}{\sqrt{g}}       }.
\end{eqnarray}
Inserting the explicit metrical functions, this miraculously is reduced to
\begin{eqnarray}
u^x=-\frac{x^2}{K}.
\end{eqnarray}
Using \eq{k1} and \eq{k2}, the general relation between $dx$ and $dx'$ is given by
\begin{eqnarray}
dx'= -\frac{1}{ {\alpha} u^x} dx.
\end{eqnarray}
In explicit form this is then reduced to
\begin{eqnarray}\label{jab}
dx'=\frac{K}{x^2 \sqrt{  \frac{ K^2 }{x^2} \left( 1-\frac{1}{x}    \right) +1     }} dx.
\end{eqnarray}
Incidentally, using the Eddington-Finkelstein original coordinates
yields the same expression, as it must. The expression however turns
out not to be particularly easy to integrate analytically except in
the limits where $K$ is either infinite or zero. In the limit where
$K$ is infinite it however cannot be inverted to find $x$ in terms of
$x'$. In any case \eq{jab} is sufficient to express the background
geometry explicitly.

\section{The background optical geometry}
Inserting (\ref{jab}) into (\ref{opmet}) we may at a {\it fix} time $t'$ write the optical line element as
\begin{eqnarray}\label{linel}
d\tilde{s}^2=\frac{1}{x^4  \left(    \frac{1}{x^2} \left(  1-\frac{1}{x}     \right)+ \frac{1}{K^2}  \right)^2}dx^2 + 
\frac{1}{   \frac{1}{x^2} \left( 1 - \frac{1}{x} \right) + \frac{1}{K^2}    }      d\Omega^2.
\end{eqnarray}
In the limit of $K\rightarrow \infty$ this takes the familiar form of the standard optical geometry
\begin{eqnarray}
d\tilde{s}^2=\frac{1}{\left(  1-\frac{1}{x}     \right)^2} dx^2 + 
\frac{x^2}{  1-\frac{1}{x} }      d\Omega^2.
\end{eqnarray}
At the other end, where $K$ goes to zero, it to lowest non-zero order approaches
\begin{eqnarray}
d\tilde{s}^2=\frac{K^4}{x^4} dx^2 + K^2 d\Omega^2.
\end{eqnarray}
Here the $x$-dependence can be taken away by another coordinate transformation. It is then obvious that in this limit we have a flat space. In a symmetry plane this would correspond to a cylinder, infinitely extended in the direction of the singularity but with finite distance from horizon to infinity. Unfortunately in the same limit the optical velocity of constant $x$ position goes to the velocity of light, to lowest order. This means that, if we just concern ourselves with the lowest order influence of $K$ on metrical components and velocities, we will not get any usable dynamics%
\footnote{Think of the rubber sheet model discussed earlier.}.
We can for instance not find the photon radius. If we still would like
to use the $K=0$ limit, we must take higher order terms into
account. Perhaps expressions when expanded to the second
non-vanishing order in $K$ will be easier to deal with than in the
general case. If this would work out it would be no approximation but
give the exactly correct dynamics. The point of considering this limit
is of course that in this limit the {\it full} spacetime is spanned by
the optical geometry. We will however not pursue this point further here.
In any case we may, for arbitrary $K$, Taylor expand $d\tilde{s}/dx$ in the limit where we approach the innermost point of the optical geometry. Doing this we readily find that regardless of $K$ the momentaneous distance to the innermost point is infinite.

Remember however that the line element of (\ref{linel}) is not strictly the optical geometry. It is not with respect to this element (except in the $K \rightarrow \infty$ limit) that photons move on geodesics, as discussed earlier. 

The speed $d\tilde{s}/dt'$ of the constant $x$ lines relative to the optical space is easy to derive since we have $K=dt'/dx'$ and from (\ref{opmet}) we see that $d\tilde{s}=K dx'/{\alpha}^2$. Then we find
\begin{eqnarray}
\frac{d\tilde{s}}{dt'} = \frac{1}{ \sqrt{ \frac{K^2}{x^2} \left(  1-\frac{1}{x}     \right)+ 1    }   }.
\end{eqnarray}
We see that in the limit where $K \rightarrow \infty$ this goes to zero as it should. At $K=0$ it goes to the velocity of light. Incidentally the velocity is at a minimum at $x=3/2$, the photon radius.

So now we have everything that we need to make explicit calculations in the generalized optical geometry, using the rubber sheet analogy. In fact we may also embed the background optical geometry and visualize the photon radius.

\section{Embedding the background geometry}
\begin{figure}[ht]
  \begin{center}
      	\epsfig{figure=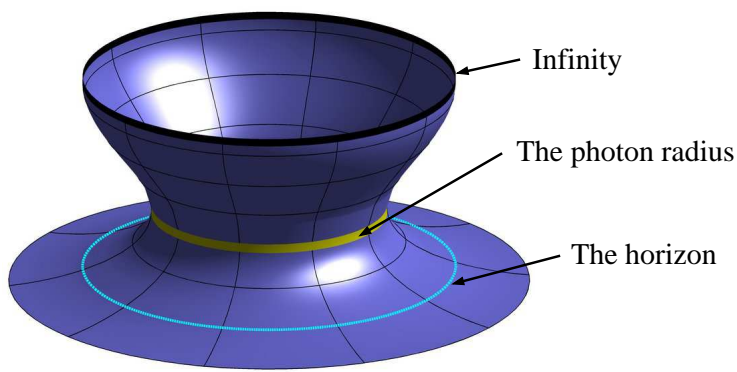,width=9.9cm,angle=0}
	\caption{The background optical geometry. The author took the liberty of enhancing the radial variations to get more shape without affecting the qualitative behavior.}  
     	\label{no1}
  \end{center} 
\end{figure}

In \fig{no1} we see a schematic picture of how an embedding of the background geometry would look. 
\noindent
Using this qualitative image we will see that the photon radius
lies exactly at the neck of the background geometry, just like in
standard optical geometry. We will also understand that the way gyroscopes in
circular motion precesses, is different inside and outside of the
neck.

\subsection{Photon geodesics}
Study now a photon moving on the surface. At any given radius we can find an angle of the velocity vector of the photon, relative to the rubber sheet, such that instantaneously the photon has no radial velocity relative to the background geometry%
\footnote{
If we direct the photon directly outwards it will move slowly out
towards spatial infinity. On the other hand if we direct it purely azimuthally relative to the rubber sheet it will be dragged inwards with the rubber sheet. Somewhere in between there is obviously an angle such that it has purely azimuthal velocity relative to the {\it background}.}.

\begin{figure}[ht]
  \begin{center}
      	\epsfig{figure=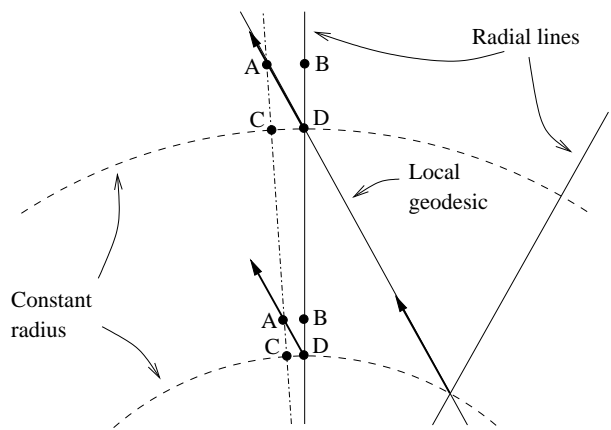,width=8cm,angle=0}
      	\caption{The shift of congruence points A,B,C and D as seen
      	from a local background geodesic coordinate system. Since the
      	congruence is non-shearing, the angle the forward direction
      	makes with a pure radial line is unaltered by the shifting of
      	the congruence points. Notice that the net effect is that the
      	forward direction is parallel transported relative to the
      	background geometry.}  
     	\label{no3}
  \end{center} 
\end{figure}

A free photon, with an initial position and velocity such that it has
no radial velocity, will follow a local geodesic on the
surface. However, its position relative to the background will be
shifted continously according to the sheet velocity. To time evolve
the the position and velocity of the photons, we may however consider
the following two-step process. First we move a distance corresponding to the time step $dt$, 
along a geodesic on the surface. Then we take the resulting forward 
direction and parallel transport it downwards corresponding
to the shift of the congruence points. In the second step, the
angle of the forward direction with a purely radial line will be
maintained. This follows from that the congruence is non-shearing (see
\fig{no3} for intuition).
We may iterate the
two-step process to time evolve the position to arbitrary times. 

\begin{figure}[ht]
  \begin{center}
      	\epsfig{figure=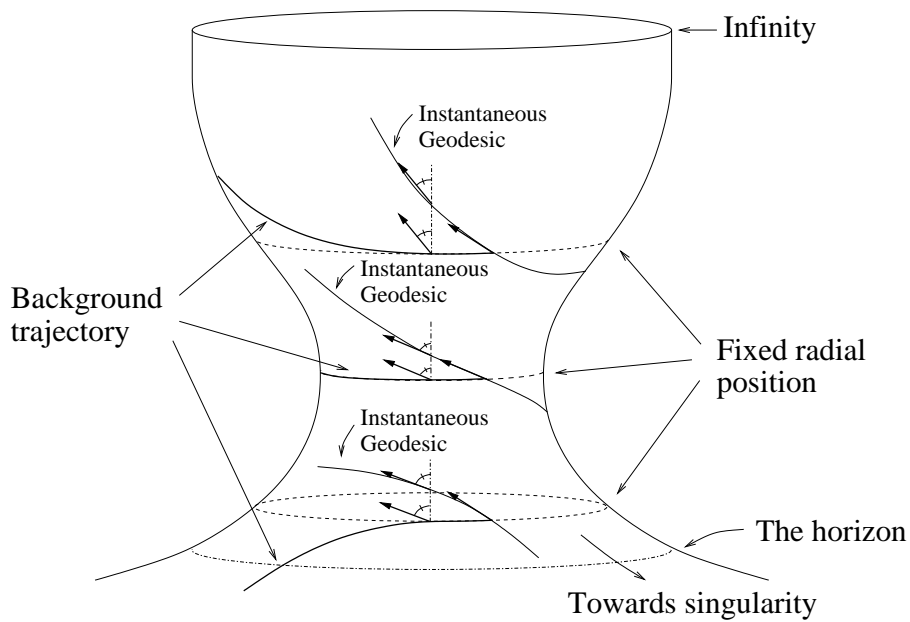,width=12cm,angle=0}
      	\caption{Understanding the photon radius in the generalized
      	optical geometry. We can time evolve the position and velocity
      	of the photon by a two-step process. First we transport the
      	forward direction along a geodesic on the surface and then we
      	parallel transport it downwards according to the shift of the
      	congruence during the time step.}  
     	\label{no2}
  \end{center} 
\end{figure}

For a geodesic on a rotational surface it is easy to show that the
angle the geodesic makes with a local line of fixed azimuthal angle (a
radial line in this case) is {\it decreasing} with increasing radius,
and vice versa. In \fig{no2} we illustrate the effects of this shift
of angles for photon geodesics with no momentary radial velocity.

\noindent

Looking at \fig{no2} we may understand that if we are on the outside of the neck, the photon velocity vector will be directed more and more outwards as time goes. Thus it will leave the radius it started at and move to infinity.
We also realize that for the corresponding initial velocity vector inside the neck, the velocity vector will be rotated to be directed less and less outwards and thus will start to move inwards. 
If we start exactly at the neck however, where the embedding radius
doesn't change to first order, the photon will remain on the same
radius. So, just like in standard optical geometry (see e.g
\cite{optiskintro}), a geodesic photon in circular motion will stay at
the neck of the embedding.

To further clarify the two-step scenario we may plot the evolution as seen from the normal of the surface. This is depicted in \fig{no3}.

\noindent

Notice that unlike the standard optical geometry (which is a subset of
this discussion) we need the velocity of the rubber sheet apart from
the background geometry to determine the paths of free photons. We
know however that for a given background geometry the sheet velocity
depends only on the embedding radius. The bigger the radius the bigger
the velocity. Also, the velocity of the sheet at infinity and at the horizon is that
of light. This is sufficient to understand the qualitative behavior
of geodesic photons.

\subsection{Gyroscope precession for circular motion}
\begin{figure}[ht]
  \begin{center}
      	\epsfig{figure=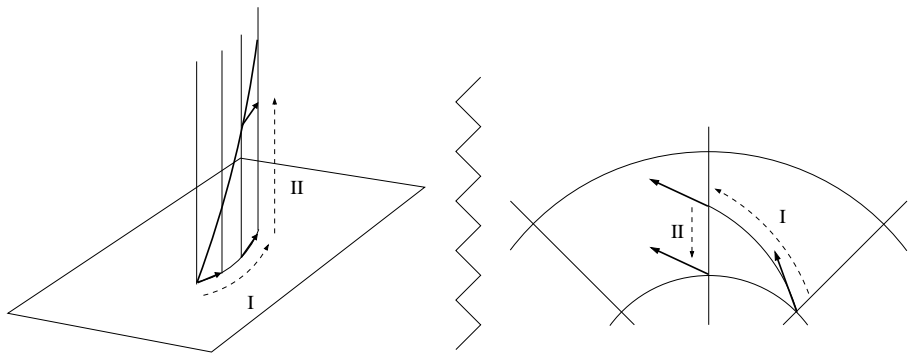,width=12cm,angle=0}
      	\caption{The two step process of moving the forward
      	direction. To the left in 2+1 dimensions, to the right in 2 dimensions.}  
     	\label{no4}
  \end{center} 
\end{figure}

Let us now consider circular motion (fixed $x$) with constant velocity. 
Consider first motion, outside the photon radius, directed to the
left as seen from outside the embedding as illustrated in \fig{no4}. 

\noindent

Moving along a circle of fixed radius with fixed speed, we {\it know} that
after the two-step process of transporting the forward direction, we
must get a forward direction that has the same angle relative to the
radial line as we had before the two-step process. As is illustrated
in \fig{no4} this means that the optical curvature has to be directed
to the left (looking at the surface from the outside).

We know from \cite{genopt} that a gyroscope undergoes pure Thomas
precession relative to the optical geometry.
For the case at hand where the trajectory turns left this means that a gyroscope will precess clockwise
relative to a corresponding parallel transported vector. 
Since the forward direction is precessing counterclockwise relative
to a parallel transported vector, the gyroscope will precess clockwise
relative to the forward direction. It follows that it will precess
clockwise relative to the local radial line also. 
Looking at the embedding from the inside (and from the top), we may say that clockwise
circular motion results in counterclockwise precession relative to the forward direction. This is in
fact what one expects of gyroscope precession in Newtonian mechanics.

Completely analogously, we may understand that inside the photon radius,
looking at the embedding from the top, clockwise circular motion
results in {\it clockwise} precession relative to the forward direction (as seen from the inside looking
at the surface). This is not analogous to the Newtonian precession.

Indeed gyroscope precession is easier to deal with in the standard
optical geometry (where the congruence is static), but as we have seen
it {\it can} be done also considering an infalling congruence, that
allows us to include the horizon.

\subsection{Inertial forces considering circular motion}
Orbiting a black hole at a fixed radius outside the photon radius requires a smaller outward
comoving force the faster one orbits the black hole (like
in Newtonian gravity). Inside the photon radius however the required
outward force increases the faster one orbits the black hole. This can be
readily understood in the standard optical geometry (see e.g
\cite{Centpar}) corresponding to infinite $K$, but when
we have a congruence moving relative to the background it is much more
complicated to see.

The point is that as we increase the orbital speed, we change the
direction of motion relative to the in-falling congruence (tilt the velocity
arrow down). This brings about all sorts of changes, for instance the
optical curvature (as opposed to the background curvature) changes. 
This particular feature of black holes apparently cannot be so easily explained,
using simple qualitative arguments, when the reference congruence is in-falling.

\section{Conclusion}
We conclude that the generalized optical geometry (assuming shearfree congruences) can be applied to
(a finite sized region of) any spherically symmetric spacetime. In
particular we can define an optical geometry from spatial infinity across the
horizon and arbitrarily close to the singularity of a static black
hole. In 2 spatial dimensions we can display the optical geometry as a
curved surface, relative to which the reference congruence points are
moving. This motion of the reference points certainly makes any
argumentation more complicated than in standard optical geometry. We
can however do
essentially the same type of qualitative arguments concerning photons,
and gyroscopes as in the standard optical geometry, and include
the horizon.
\\
\\
{\bf References}
\\

\end{document}